\documentclass[runningheads,a4paper]{aa}

\usepackage{graphicx,url,twoopt,natbib}
\usepackage{multirow}
\usepackage[varg]{txfonts}
\usepackage{hyperref}
\usepackage{xcolor}
\usepackage{linenoaa}

\hypersetup{
  colorlinks=true,
  citecolor=blue,
  urlcolor=blue,
  linkcolor=blue
}

\bibpunct{(}{)}{;}{a}{}{,}
\makeatletter
  \protected\def\stonyslink{
     \def\hyper@linkstart##1##2{}\let\hyper@linkend\@empty}
  \newcommandtwoopt{\citeads}[3][][]{
   \href{http://ui.adsabs.harvard.edu/abs/#3/abstract}
        {\stonyslink \citealp[#1][#2]{#3}}
   \biblink{#3}{\href{http://ui.adsabs.harvard.edu/abs/#3/abstract}{ADS}}}
 \newcommandtwoopt{\citepads}[3][][]{
   \href{http://ui.adsabs.harvard.edu/abs/#3/abstract}
        {\stonyslink \citep[#1][#2]{#3}}
   \biblink{#3}{\href{http://ui.adsabs.harvard.edu/abs/#3/abstract}{ADS}}}
 \newcommandtwoopt{\citetads}[3][][]{
   \href{http://ui.adsabs.harvard.edu/abs/#3/abstract}
        {\stonyslink \citet[#1][#2]{#3}}
  \biblink{#3}{\href{http://ui.adsabs.harvard.edu/abs/#3/abstract}{ADS}}}
 \newcommandtwoopt{\citeyearads}[3][][]{
   \href{http://ui.adsabs.harvard.edu/abs/#3/abstract}
        {\stonyslink \citeyear[#1][#2]{#3}}
   \biblink{#3}{\href{http://ui.adsabs.harvard.edu/abs/#3/abstract}{ADS}}}
\makeatother

\begin{document} 

   \title{Spectral properties of mesoscale fluctuations in interplanetary coronal mass ejections}

      \author{Anna-Sofia Jylh\"{a}
          \inst{1,2}
          \and Simon Good
          \inst{1,}\thanks{Corresponding author: simon.good@helsinki.fi}
          \and Emilia Kilpua
          \inst{1}  
          }

     \institute{Department of Physics, University of Helsinki, PO Box 64, FI-00014 Helsinki, Finland
   \and Department of Physics, University of Alberta, Edmonton AB T6G 2E1, Canada \\
   }
                 
   \date{Received 19 July 2026 / Accepted 27 October 2026}

  \abstract{Magnetic fields in interplanetary coronal mass ejections (ICMEs) often display a flux rope structure at large scales and a turbulent cascade of fluctuations at smaller scales. However, the nature of fluctuations at transition scales between the flux rope and turbulence -- i.e., at mesoscales -- has not been previously examined in detail. Mesoscale fluctuations of the magnetic field in 142 ICME intervals with clear flux rope signatures at 1~au have been examined using wavelet power spectra, with ``mesoscale'' defined here as spacecraft-frame frequencies $10^{-4}\ \text{Hz}<f_\mathrm{sc}< 10^{-3}\ \text{Hz}$, equivalent to timescales $2.8\ \text{hr}\gtrsim \tau \gtrsim 17\ \text{min}$. The spectra were found to be significantly less steep at mesoscales than at larger and smaller scales, with an average spectral slope close to $-1$ at these intermediate scales. The mesoscale fluctuations had relatively high cross helicity magnitudes and low compressibility, suggesting a significant degree of Alfv\'enicity, while also having more negative residual energy than the turbulent fluctuations found at smaller scales. Uncertainties in the spectral slope values at mesoscales were used to distinguish between spectra showing a relatively smooth power-law behavior or (less commonly) a more bumpy trend; the smoother power-law intervals had a balanced cross helicity distribution while the more bumpy spectra had a predominantly antisunward cross helicity. Slow wind intervals were similarly analyzed, and showed similar properties to the ICMEs. These results indicate that a spectrally narrow $1/f$ range is typically present in ICMEs at mesoscales.}
  

   \maketitle
    \nolinenumbers
\section{Introduction}

The solar wind is a highly complex plasma environment that fluctuates at all measurable scales. These fluctuations represent a mix of structures, turbulence and waves in the plasma, with the nature of this mix varying from scale to scale \citep[e.g.][]{Verscharen19}. At large scales, the magnetic structure of the solar wind is reasonably well described by the Parker spiral \citep{Parker58}. Superimposed on this background structure at scales exceeding the correlation length are large-amplitude Alfv\'en waves with an antisunward sense of propagation and a $1/f$ scaling of power spectral density \citep{BelcherDavis71,DenskatNeubauer82,Bavassano82,MatthaeusGoldstein82}. This $1/f$ or ``injection'' range is a well established feature of the fast wind but has also recently been identified more systematically in slow wind \citep{Bruno19,Dorseth24a,Dorseth24b}. The origin of the $1/f$ range is much debated, with theories based, for example, on wave generation by photospheric or chromospheric motions \citep[e.g.][]{Velli89,Cranmer05,Verdini12}, a superposition of uncorrelated samples of solar-surface turbulence \citep{MatthaeusGoldstein86}, or limitations imposed on the spectral scaling by incompressibility \citep{Matteini18}. At scales smaller than the correlation length, the magnetic field spectral index rolls over to values that are are usually close to either $-3/2$ or $-5/3$, these indices being consistent with theories of Alfv\'enic magnetohydrodynamic (MHD) turbulence \citep[e.g.][]{Chen16}. In the canonical turbulence picture, energy supplied from the injection range is cascaded to successively smaller scales by the MHD turbulence and ultimately dissipated at ion and electron scales \citep[e.g.][]{Alexandrova09}.

Interplanetary coronal mass ejections (ICMEs) provide an environment in which to explore this multiscale variation of fluctuation properties that is different to the fast and slow winds. For instance, in contrast to the more continuous solar-wind outflow, ICMEs are transient heliospheric phenomena with magnetic drivers that often have a flux rope structure at large scales \citep{Goldstein83}. A spacecraft will observe the passage of an ICME flux rope in situ as a large-amplitude rotation of the magnetic field vector. \citet{Good23,Good26} found that this rotating background field has a significant steepening effect on spectral slopes at the lowest spacecraft-frame frequencies, $f_{\textrm{sc}}$, that can be sampled within the ICME driver intervals. Slope values at these flux rope scales (typically $f_{\textrm{sc}}\lesssim10^{-4}$~Hz at 1~au) are often $-2$ or steeper, rather than the $-1$ value found in fast or slow solar wind at the same scales. Magnetic field fluctuations at $f_{\textrm{sc}}\gtrsim10^{-3}$~Hz typically have a spectral scaling closer to $-5/3$ than $-3/2$ (\citealp[e.g.,][]{Liu06,Borovsky19,Good23,Shaikh24}; \citealp[cf.][]{Hamilton08}), likely due to the low average values of cross helicity found in ICME drivers \citep{Roberts20,Good20,Good22,Soljento23}.

In ICMEs, the spectral properties of magnetic field fluctuations at scales between the global flux-rope scale and the fully developed inertial range of the MHD turbulence have not been examined in detail previously. These transitional scales, at $10^{-4}\ \text{Hz}\lesssim f_{\textrm{sc}}\lesssim10^{-3}\ \text{Hz}$ (i.e. timescales from $\sim$17~min to $\sim$2.8~hr) at 1~au, fall within the scope of the variously defined mesoscales that are of increasing interest in ICME research \citep{Lugaz18,Viall21,Palmerio24,Liu24,Scolini24,Scolini25}. While ``mesoscale'' is sometimes used to refer to ICME substructure (e.g. the compressed sheaths formed immediately upstream of drivers), it can also encompass complex or non-ideal features in the driver intervals that deviate from simple flux rope models. Applying Taylor's hypothesis such that wavenumber $k=2\pi f_\textrm{sc}/v_{\textrm{sw}}$ is related to length scale $l$ as $k=1/l$, the $10^{-4}\ \text{Hz}< f_{\textrm{sc}}<10^{-3}\ \text{Hz}$ mesoscale frequencies we have analyzed correspond to $64~\textrm{Mm}\lesssim l \lesssim640~\textrm{Mm}$ for a bulk flow speed $v_{\textrm{sw}}=400~\textrm{km~s}^{-1}$. We note that these scales are an order of magnitude smaller than the 0.005--0.05~au (i.e. 750--7500~Mm) mesoscale range defined by \citet{Lugaz18}, but fall within the 5--10\,000~Mm range defined by \citet{Viall21}.

Mesoscale features in ICMEs are often identified and characterized in the time domain, i.e., directly from time series observations made by spacecraft. In this study, we consider the frequency-domain properties of mesoscale fluctuations in a large statistical sample of ICME drivers observed at 1~au; here we exclude non-driver spatial substructures such as sheaths from our definition of ``mesoscale''. The drivers were selected on the basis of having clear flux rope signatures at large scales, with the explicit aim of examining mesoscale behavior in non-complex ICMEs. Rather than a simple roll-over from the steepened spectral index associated with the flux rope field at large scales to the approximately $-5/3$ index of the MHD turbulence at small scales, we have found a third, distinct band of fluctuations to be typically present at mesoscales in these ICMEs, often with a spectral index close to $-1$. A possible interpretation of this band is that is corresponds to the high-frequency end of an injection range similar to that found in other solar wind types, with the coherent flux rope rotation in ICMEs obscuring or suppressing the range at lower frequencies. Rather than being a dynamic injection range that supplies energy to the turbulent cascade, the mesoscale band could also comprise relatively static magnetic structure. Both of these possibilities are considered and discussed.

In Section~\ref{sect:spacecraft_data}, the spacecraft data and ICME intervals selected for the analysis are described. In Section~\ref{sect:analysis}, a statistical analysis of ICME mesoscale fluctuations is presented. Key parameters analyzed include power spectral density, spectral index, correlation times, residual energy, cross helicity, magnetic helicity, and magnetic compressibility. The results of this analysis are compared and contrasted with a similar analysis of slow wind intervals observed at 1~au. In Section~\ref{sect:discussion}, the broader implications and significance of our findings are discussed.

\section{Spacecraft data} \label{sect:spacecraft_data}

Data from the \textit{Wind} spacecraft between 1995 and 2020 were analyzed. A total of 142 ICMEs with clear flux rope signatures have been selected from NASA's ICME catalog \citep{Nieves18}, with magnetic drivers (``obstacles'') classified in the catalog as having a well-ordered, monotonic field rotation of 90$^{\circ}$--180$^{\circ}$ (``Fr''), less than 90$^{\circ}$ (``Fr-''), or more than 180$^{\circ}$ (``Fr+''). Only the magnetic driver intervals of the ICMEs have been analyzed and not other ICME substructures, such as sheaths; all mentions of ICMEs in this work refer to the magnetic drivers only. The 142 events were also selected on the basis of having a duration of at least 15 hours and having no significant data gaps (i.e. $<5\%$). The selected ICMEs represent 36\% of all ICMEs listed by the catalog during 1995-2020, with the remainder comprising complex drivers (``C''), non-flux rope ejecta (``E''), or flux rope intervals not meeting the minimum duration or data gap selection criteria. Twenty-six slow solar wind intervals of 24-hr duration with proton radial speeds below 400 km s$^{-1}$ throughout the interval have been selected to provide an auxiliary study. Magnetic field measurements by MFI \citep{Lepping95} and ion moments by 3DP/PESA-L \citep{Lin95} have been used at 3~s and 1~min average resolution, respectively.

\section{Analysis} \label{sect:analysis}

\subsection{Magnetic field power spectra} \label{sect:individual_psds}

The Morlet wavelet technique \citep[e.g.][]{Torrence98} has been used to calculate wavelet spectrograms of the trace power spectral density (PSD) of the magnetic field, $E_B$, such that $E_B=2\pi\sum_i|\mathcal{W}_{B_i}|^2/f_\mathrm{sc}$, where $\mathcal{W}_{B_i}(s,t)$ are the wavelet amplitudes of magnetic field components $B_i$ in geocentric solar ecliptic (GSE) coordinates $i=x, y, z$, and where wavelet scale $s\simeq1/f_\mathrm{sc}$; further details of the wavelet analysis are found at the IRFU-Matlab repository \citep{Khotyaintsev24}. The $E_B$ spectrograms have been time-averaged to obtain the mean PSDs in the 142 flux rope intervals as a function of frequency. Spectral indices, $\alpha$, were calculated from these mean PSDs over a sliding frequency window with a logarithmically fixed width of 0.7 decades. Significantly shorter window lengths did not give statistically valid results, while significantly longer window lengths were too broad to show features in relatively narrow frequency ranges of interest in the power spectra, i.e., the transition between the flux rope scale and the inertial range; windows with widths similar to 0.7 (e.g. ~0.5--1) provide comparable results to the ones presented here. 

Spectral indices of the magnetic field PSD calculated with a sliding window can be used to characterize the behavior of the power spectrum at different scales. Examination of the sliding-window $\alpha$ profiles in the frequency range $10^{-4}\ \text{Hz}<f_\mathrm{sc}< 10^{-3}\ \text{Hz}$ (our definition of ``mesoscale'') reveals a common trend: a local maximum in the signed $\alpha$ value, representing a flattening of the PSD relative to the steeper spectrum (i.e. lower values of signed $\alpha$) found either side of the mesoscale band. The 142 ICMEs have been categorized in terms of the presence of this feature, with 96 of the 142 having a robustly identified $\alpha$ maximum in the mesoscale band.
The other 46 events had local maxima that fell outside the mesoscale band or that were less clearly defined, or did not have maxima. The 96-event subset has been analyzed in further detail.

   \begin{figure*}
   \centering
   \includegraphics[width=0.9\textwidth]{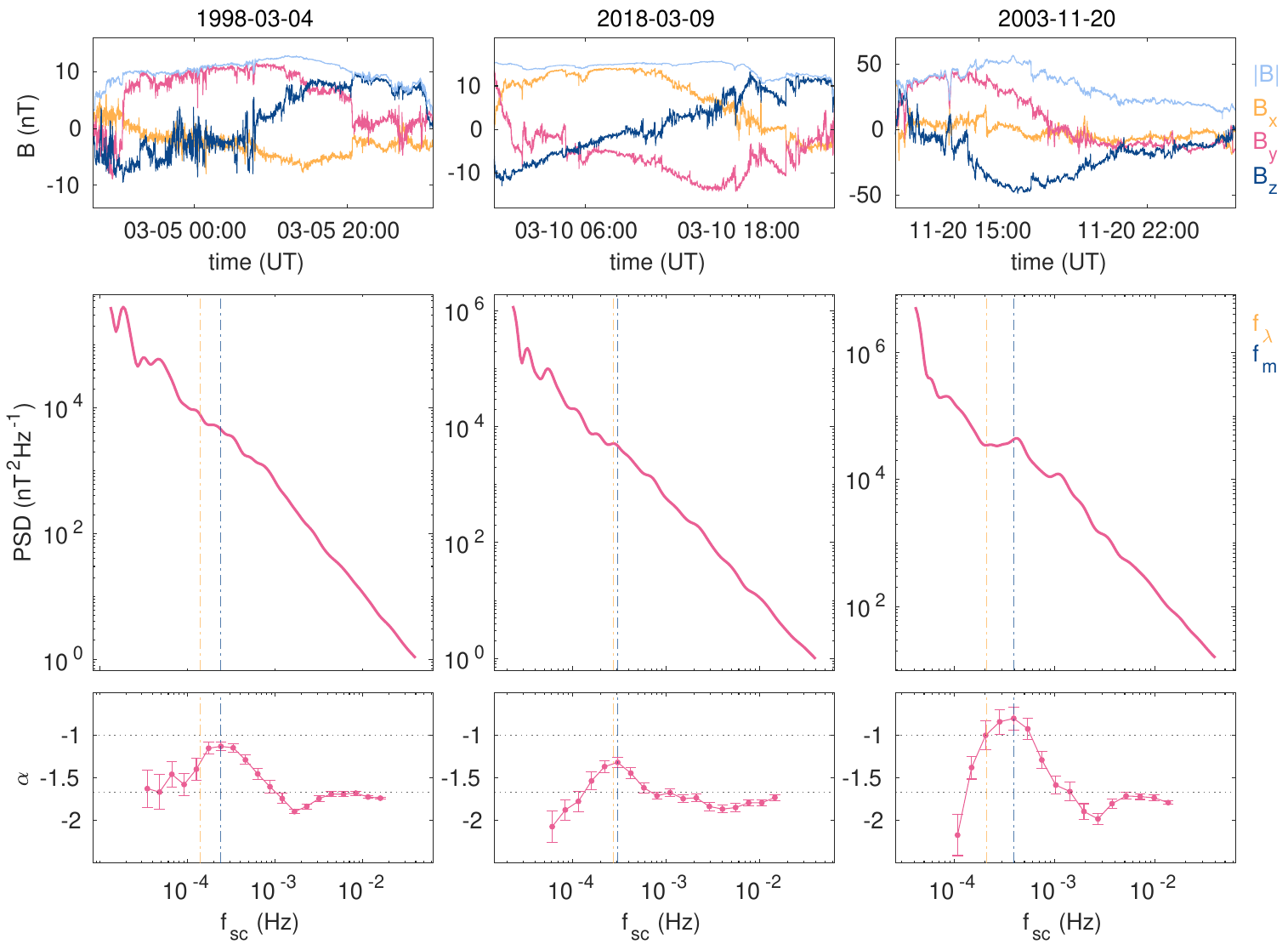}
   \caption{Time-averaged wavelet PSD of the magnetic field in three example ICMEs. From top to bottom, the panels show the magnetic field magnitude and components in GSE coordinates, the PSD, and the spectral index calculated with a sliding window. Vertical lines display the correlation frequency, $f_\lambda$, and the frequency of the largest signed $\alpha$ value in the mesoscale band, $f_\mathrm{m}$.}
    \label{fig:CME_event_examples}
    \end{figure*}

Three example events are shown in Fig.~\ref{fig:CME_event_examples}. The top panels show the magnetic field time series, the middle the time-averaged PSD and the bottom the sliding-window $\alpha$ values. The frequency of the local maximum in the $\alpha$ values is shown with a dark blue dash-dotted line. The apricot dash-dotted line shows the correlation frequency $f_\lambda=1/\lambda$, where the correlation time, $\lambda$, is defined as the point where the autocorrelation curve $C(\tau)$ equals $1/\mathrm{e}$. The autocorrelation curve has here been calculated as
    \begin{equation}
        C(\tau) = \frac{4\langle\delta \vec{B}(t)\cdot\delta\vec{B}(t+\tau)\rangle}{\langle(\lvert\delta \vec{B}(t)\rvert+\lvert\delta\vec{B}(t+\tau)\rvert)^2\rangle},
        \label{eqn:autocorrelation curve}
    \end{equation}
where $\tau$ is a time lag and $\delta\vec{B}(t) = \vec{B}(t) - \vec{B}_\mathrm{FR}(t)$, with $\vec{B}_\mathrm{FR}(t)$ representing the flux rope magnetic field estimation at a given time, obtained by fitting a second-order polynomial to each magnetic field component. The flux rope field was subtracted in order to estimate the correlation time of the fluctuations rather than the background flux rope, as detailed by \citet{Good26}. While lacking the physical constraints of standard flux rope models \citep[e.g. the Lundquist model;][]{Lepping90}, polynomial fitting provides a sufficiently accurate detrending method for the purpose of estimating correlation lengths in a large sample of events that are not all well fitted by the standard models.

Local maxima in the signed $\alpha$ values are clearly visible at scales similar to the correlation frequencies in the three events in Fig.~\ref{fig:CME_event_examples}. Despite the general resemblance in the three $\alpha$ profiles, the power spectra at mesoscales differ somewhat between the events. Of the three example events, the first two show a relatively smooth power-law behavior (i.e. linear on a log-log plot) while the third is more bumpy at mesoscales. In cases with significant bumps in the power spectrum, the uncertainties of the fits are larger; this aspect is examined in detail in Section~\ref{sect:uncertainties}.

    \begin{figure}
   \centering
   \includegraphics[width=0.49\textwidth]{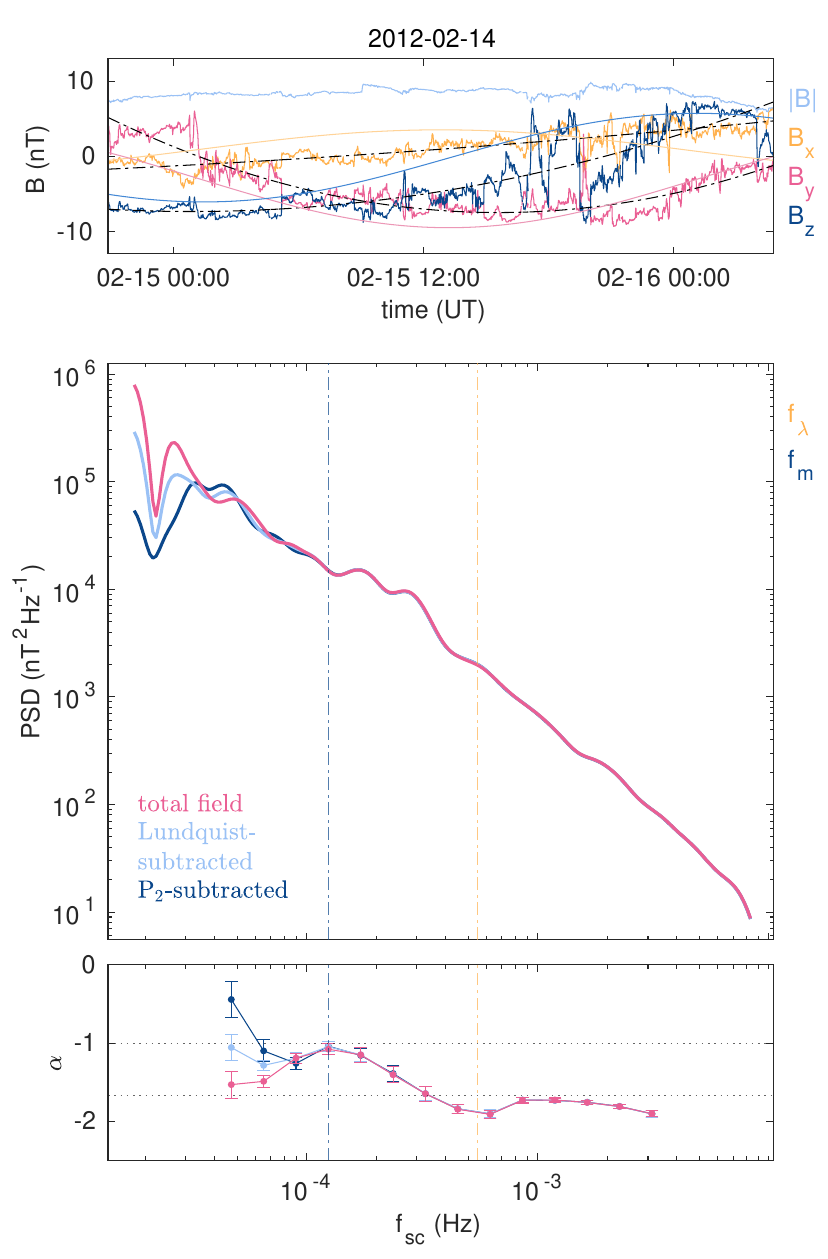}
   \caption{Effect of flux rope field subtraction on the magnetic field PSD. The top panel shows the magnetic field magnitude and components in GSE coordinates with fits to each component. The Lundquist fit is shown by the colored lines and second-order polynomial fits by the black dash-dotted lines. The middle panel shows the magnetic field PSD for the total field, with Lundquist fit subtraction, and with polynomial subtraction. The bottom panel shows the sliding window spectral indices for the three PSDs.}
    \label{fig:CME_fits_example_event}
    \end{figure}

An estimation of the flux rope field can also be removed from the data before calculating the PSD, resulting in a less steep spectrum at the largest scales \citep{Good23,Good26}. However, the mesoscales are at most weakly affected by the flux rope field being removed from the data. This is illustrated with an example event in Fig.~\ref{fig:CME_fits_example_event} where the flux rope field was estimated with a force-free Lundquist fit to the data \citep{Lepping90,Good19} and also with second-order polynomial fits to the magnetic field components. The fits and the corresponding magnetic field time series are shown in the top panel, with black dash-dotted lines denoting the polynomial fit and colored lines the Lundquist fit. The second and third panels show the PSD and $\alpha$ values for the total field as well as the Lundquist-subtracted and polynomial-subtracted cases. The correlation frequency and maximum $\alpha$ value at mesoscales are indicated with apricot and blue dash-dotted lines, respectively. It is evident in Fig.~\ref{fig:CME_fits_example_event} (and generally true for the other events) that the mesoscales are well separated from the global flux rope scale and are largely unaffected by the flux-rope field subtraction. We have thus opted to analyze mesoscale spectral properties without any flux rope subtraction.

\begin{figure*}
   \centering
   \includegraphics[width=1\textwidth]{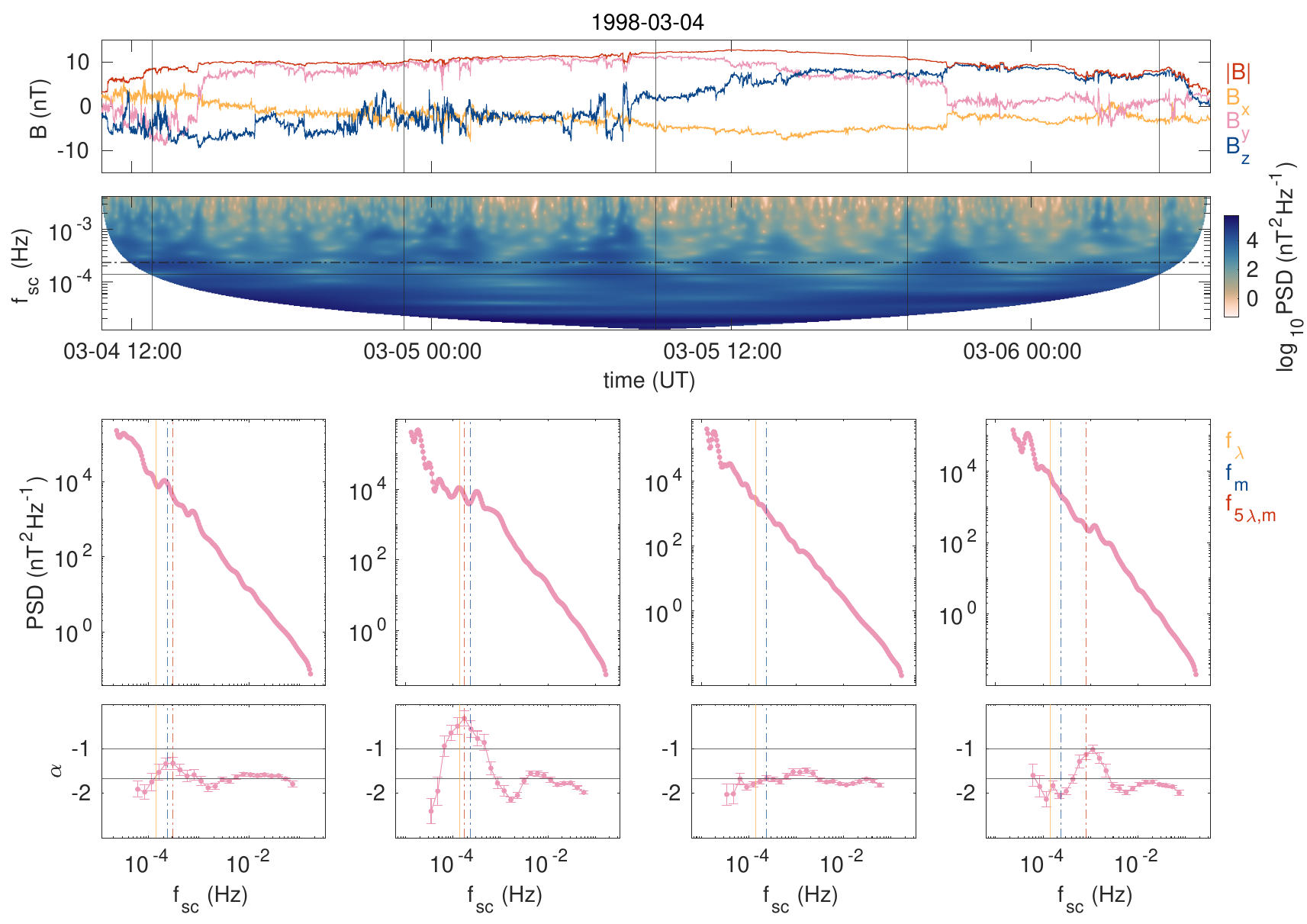}
   \caption{An example of $5\lambda$ sub-interval analysis. The top two panels shows the magnetic field time series and corresponding PSD spectrogram, with sub-intervals marked by vertical lines. The two bottom rows show the time-averaged PSD of each sub-interval and corresponding sliding-window $\alpha$ profile. Vertical lines in the bottom rows denote the correlation frequency, $f_\lambda$, and frequency of the largest $\alpha$ at mesoscales for the entire interval, $f_{\mathrm{m}}$, and for each sub-interval, $f_{5\lambda,\mathrm{m}}$.}
    \label{fig:CME_subintervals_example_event}
    \end{figure*}

The wavelet spectrograms have also been time-averaged over shorter sub-intervals to sample properties within the ICMEs more locally. Sub-intervals with durations equal to five correlation times ($5\lambda$) in each ICME were chosen, giving intervals short enough for local sampling and long enough to robustly resolve the mesoscale fluctuations (i.e. $5\lambda\gg1/f_\mathrm{m}$). The time remaining after partitioning the total time into $5\lambda$ intervals was divided equally between the start and end of the event, and the portions of the wavelet spectrograms associated with these times were left out of the analysis. Sub-intervals with more than 10\% of PSD values at frequencies higher than $f_\lambda$ or $10^{-4}$~Hz (whichever is lower) lying outside the cone of influence were also discarded. Fig.~\ref{fig:CME_subintervals_example_event} shows an example where the spectra has been divided into $5\lambda$ intervals. The top panel shows the magnetic field time series, with the corresponding spectrogram shown immediately below. The sub-intervals have been marked with vertical lines. The bottom two rows respectively show the time-averaged PSD and the sliding-window $\alpha$ for each $5\lambda$ sub-interval. The correlation frequency for the full interval is marked with the apricot lines, and the frequencies of the local signed maxima of the $\alpha$ values for the entire event and each sub-interval are shown by blue and red dash-dotted lines, respectively. The sub-intervals differ in the value of the maximum signed $\alpha$ and the scale at which this $\alpha$ occurs; some sub-intervals have relatively smooth spectra at mesoscales, while others are more bumpy. The mesoscale band in the third interval does not show any distinguishing properties, having the same slope as the turbulence at smaller scales. Across all the events analyzed, mesoscale properties of the $5\lambda$ sub-intervals were not found to vary systematically across the full intervals; for example, there were no statistical differences between sub-intervals sampling the front, middle, or rear parts of the ICMEs.

\subsection{Statistical analysis} \label{sect:statistical_analysis}

\begin{figure}
   \centering
   \includegraphics[width=0.48\textwidth]{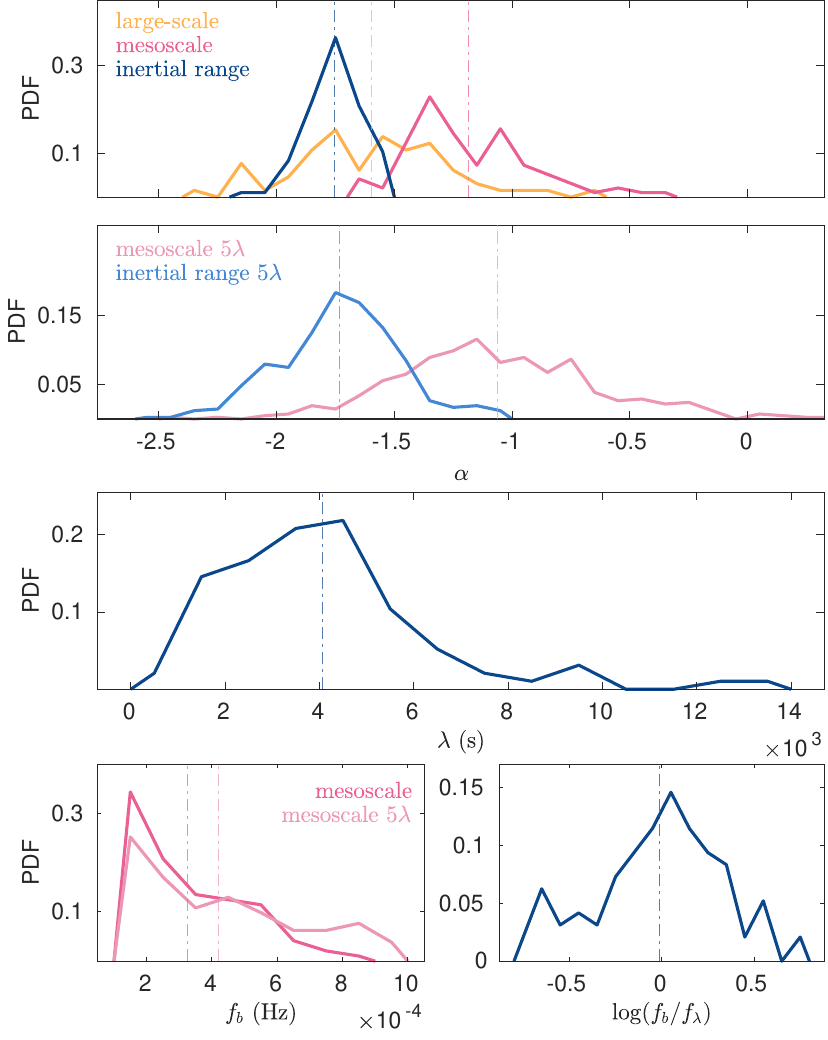} 
   \caption{PDFs of $\alpha$ values and other key parameters. The figure shows: the maximum $\alpha$ value for the mesoscales, the mean $\alpha$ values in the inertial range, and mean values in the large-scale range (top panel); the equivalent $\alpha$ values in the mesoscale and inertial ranges for the $5\lambda$ sub-intervals (second panel); the $\lambda$ values (third panel); the frequency at which the largest (i.e. shallowest) $\alpha$ value is found at mesoscales in the full intervals and $5\lambda$ sub-intervals (bottom left panel); and the logarithmic ratio of the frequency associated with the largest mesoscale $\alpha$ value and correlation frequency (bottom right panel). Vertical lines mark the distribution means.}
    \label{fig:CME_alpha_histograms}
    \end{figure}

The statistics of the maximum signed $\alpha$ at mesoscales across multiple events are usefully represented with distributions. These are displayed in Fig.~\ref{fig:CME_alpha_histograms}, where the means of the probability density functions (PDFs) are shown with dash-dotted lines of the same color as the corresponding distribution lines. The top panel of Fig.~\ref{fig:CME_alpha_histograms} shows $\alpha$ values at three scales: the low frequency end of the inertial range ($10^{-3}$--$10^{-2}$~Hz), larger scales (<\ $10^{-4}$~Hz), and mesoscales ($10^{-4}$--$10^{-3}$~Hz). The mesoscale $\alpha$ values shown in the distributions are the maximum value in the mesoscale range, while a mean over the whole range was taken for the other two. The mean and standard deviation of the resulting distributions are, respectively, $-1.75$ and 0.12 for the inertial range, $-1.19$ and 0.27 for the mesoscales, and $-1.60$ and 0.33 for the larger scales. The large-scale distribution comes from the analysis of a subset ($n_\alpha=65$) of the full 96-event
list that excludes events too short in duration for obtaining accurate $\alpha$ values at $f_\textrm{sc}<10^{-4}$~Hz. Throughout this work, statements comparing distribution means have been tested for their statistical significance using Welch's t-test; unless otherwise stated, statements are supported to at least a 5\% significance level.

\begin{table*}[ht]
\caption{Key parameters obtained from the spectral analysis of the ICMEs and slow solar wind intervals.}
\label{tab:key_parameters}
\centering
\begin{tabular}{l c c c c c c c c c c c}
\hline\hline
 & $\alpha$ & $s_\alpha$ & $n_\alpha$ & $\sigma_\mathrm{r}$ & $s_{\sigma_\mathrm{r}}$ & $\sigma_\mathrm{c}$ & $s_{\sigma_\mathrm{c}}$ & $|\sigma_\mathrm{m}|$ & $s_{|\sigma_\mathrm{m}|}$ & $c$ & $s_c$ \\
\hline
\\
\textit{ICMEs} \\
Inertial range (5$\lambda$) & $-1.73$ & 0.25 & 416 & $-0.35$ & 0.15 & 0.02 & 0.28 & 0.06 & 0.05 & 0.04 & 0.04 \\
Mesoscale (5$\lambda$) & $-1.06$ & 0.42 & 416 & $-0.45$ & 0.25 & $-0.01$ & 0.42 & 0.17 & 0.15 & 0.05 & 0.06 \\
Large scale & $-1.60$ & 0.33 & 65 & $-0.51$ & 0.24 & 0.01 & 0.25 & 0.28 & 0.09 & 0.10 & 0.08 \\
\\
\textit{Slow solar wind} \\
Inertial range (5$\lambda$) & $-1.59$ & 0.25 & 106 & $-0.36$ & 0.12 & 0.20 & 0.32 & 0.06 & 0.03 & 0.09 & 0.07 \\
Mesoscale (5$\lambda$) & $-1.16$ & 0.39 & 106 & $-0.49$ & 0.19 & 0.24 & 0.51 & 0.14 & 0.14 & 0.05 & 0.07 \\
Large scale & $-1.39$ & 0.42 & 26 & $-0.44$ & 0.17 & 0.20 & 0.38 & 0.26 & 0.09 & 0.08 & 0.07 \\
\hline
\end{tabular}
\tablefoot{Parameters listed in the table include mean values of the trace magnetic field PSD spectral slope, $\alpha$, residual energy, $\sigma_\mathrm{r}$, rectified cross helicity, $\sigma_\mathrm{c}$, absolute magnetic helicity, $|\sigma_\mathrm{m}|$, and magnetic compressibility, $c$. The standard deviation of parameter $X$ is denoted by $s_X$. The number of intervals used to calculate the mean slope values, $n_\alpha$, is also listed.}
\end{table*}

The large-scale mean is significantly lower than $-1$ but does not reach the $-2$ or steeper value expected for the flux rope, as the window is too large (i.e. extends to too high frequencies) to capture the slope of the flux rope field in isolation. The mesoscale mean value is closer to $-1$ than the steeper values associated with a developed turbulent cascade. The distributions together show that the relative shallowness of the PSD that is observed in individual events at mesoscales is also present statistically, as the $\alpha$ distribution for the mesoscales is shifted to higher signed values than the distributions for the larger and smaller scales.

Analogously to the top panel, the second panel in Fig.~\ref{fig:CME_alpha_histograms} shows the $5\lambda$ sub-interval distributions. The sub-intervals at the large scale have been omitted because they are strongly affected by edge data gaps in the spectrogram associated with the cone of influence. Although the sub-interval distributions are wider, they are otherwise very similar to the corresponding whole-event distributions, with the mesoscale sub-intervals having slightly larger $\alpha$ values; the means and standard deviations for the distributions are $-1.73$ and 0.25 for the inertial range and $-1.06$ and 0.42 for the mesoscales. These and other key parameters obtained from the statistical analysis are listed in Table~\ref{tab:key_parameters}.

The third panel in Fig.~\ref{fig:CME_alpha_histograms} shows the correlation times for each event as calculated using Eq.~\ref{eqn:autocorrelation curve}. They vary over a range of nearly 1.5 decades; the mean and standard deviation of the distribution are $4.07\times10^{3}$~s and $2.32\times10^{3}$~s, respectively. Similarly, the left bottom-row panel shows the variation of the frequency, $f_b$, at which the largest signed $\alpha$ value is found in the mesoscale band, for both the full events and $5\lambda$ sub-intervals. The peak of the distribution is weighted toward lower frequencies within the mesoscale band. The means and standard deviations are, respectively, $3.24\times10^{-4}$~Hz and $1.73\times10^{-4}$~Hz for the entire-event distributions, and $4.19\times10^{-4}$~Hz and $2.48\times10^{-4}$~Hz for the $5\lambda$ sub-interval distributions. 

The frequencies associated with the largest signed $\alpha$ values at mesoscales and the correlation frequencies were compared by taking their ratio. The distribution of the logarithm of this ratio is shown in the bottom right panel of Fig.~\ref{fig:CME_alpha_histograms}. As both $f_b$ and the correlation frequencies vary over a large range of values, logarithmic values were taken in order to minimize the effect of outliers. The resulting distribution of $\log(f_\mathrm{b}/f_\lambda)$ has a mean value of $-0.011$ and standard deviation of 0.34, indicating that the two frequencies are typically similar in value but sometimes differ significantly.

To confirm that the relatively shallow mesoscale slopes are physical in origin, we repeated our analysis with a synthetic control dataset. This dataset comprised time series generated via inverse Fourier transforms applied to double power-law spectra with a $-2$ slope at large scales mimicking the flux rope, a $-5/3$ slope at small scales mimicking the turbulence, and a break between the two located in the middle of the mesoscale band (i.e., spectra without a $-1$ slope at mesoscales). Via the assignment of random phases, 100 time series were generated from a spectrum with a sharp break and 100 from a spectrum with a smooth break. The full-interval and $5\lambda$-interval wavelet analysis pipeline was then applied to this dataset. In both cases, for both the sharp and smooth-break datasets, the resulting distributions of mesoscale $\alpha$ (equivalent to the Fig.~\ref{fig:CME_alpha_histograms} distributions; not shown) peaked sharply at values between $-2$ and $-5/3$, with only a negligible part of the distributions falling near $-1$. This analysis confirms that the mesoscale shallowness in the real dataset is due to the presence of physically distinctive fluctuations rather than, for example, spectral leakage from larger scales in our wavelet analysis (complementing the Fig.~\ref{fig:CME_fits_example_event} analysis) or simply arising from the break in an underlying double power-law spectrum.

\subsection{Fluctuation Alfv\'enicity}

While the power spectra and spectral indices reveal a relative shallowness at mesoscales, they do not reveal the nature of the mesoscale fluctuations more generally, e.g., the extent to which they are Alfv\'enic or non-Alfv\'enic. Trace spectral densities for proton velocity, $\vec{v}$, the Alfv\'en normalized magnetic field, $\vec{b}$, and the Elsasser variables, $\vec{z^\pm}=\vec{v}\pm\vec{b}$, denoted as $E_v$, $E_b$ and $E_\pm$ respectively, were used to obtain the normalized residual energy and cross helicity. The trace PSDs of these parameters were obtained in the same way as for $E_B$, using the wavelet analysis technique described in Section~\ref{sect:individual_psds}. The Alfv\'en normalized magnetic field is defined as $\vec{b}=\vec{B}/\sqrt{\mu_0\rho}$, where $\rho$ is the ion mass density, assumed to consist of 4\% alpha particles and 96\% protons such that $\rho=7m_\mathrm{p}n_\mathrm{p}/6$, with $m_\mathrm{p}$ and $n_\mathrm{p}$ denoting the proton mass and number density, respectively. The normalized residual energy is defined as
    \begin{equation}
        \sigma_\mathrm{r} = \frac{E_v-E_b}{E_v+E_b}
        \label{eqn:sigma_r}
    \end{equation}
and the normalized cross helicity as
\begin{equation}
        \sigma_\mathrm{c} = \frac{E_+-E_-}{E_++E_-}.
        \label{eqn:sigma_c}
    \end{equation}
The cross helicity was rectified so that positive and negative values respectively indicate an antisunward and sunward sense of propagation relative to the equivalent Parker spiral field at 1~au \citep[e.g.][]{Good22}. Also calculated were the normalized magnetic helicity,
    \begin{equation}
        \sigma_\mathrm{m} = \frac{2\operatorname{Im}(\mathcal{W}_{B_y}^*\cdot \mathcal{W}_{B_z})}{|\mathcal{W}_{B_x}|^2+|\mathcal{W}_{B_y}|^2+|\mathcal{W}_{B_z}|^2},
        \label{eqn:sigma_m}
    \end{equation}
and magnetic compressibility, defined here as
    \begin{equation}
        c = \frac{E_{\lvert B \rvert}}{E_B},
        \label{eqn:compressibility}
    \end{equation}
where the imaginary part is taken in the numerator of Eq.~\ref{eqn:sigma_m}, where $\mathcal{W}_{B_y}^*$ is the complex conjugate of $\mathcal{W}_{B_y}$, and where $E_{\lvert B \rvert}$ is the PSD of the magnetic field magnitude.

\begin{figure}
   \centering
   \includegraphics[width=0.49\textwidth]{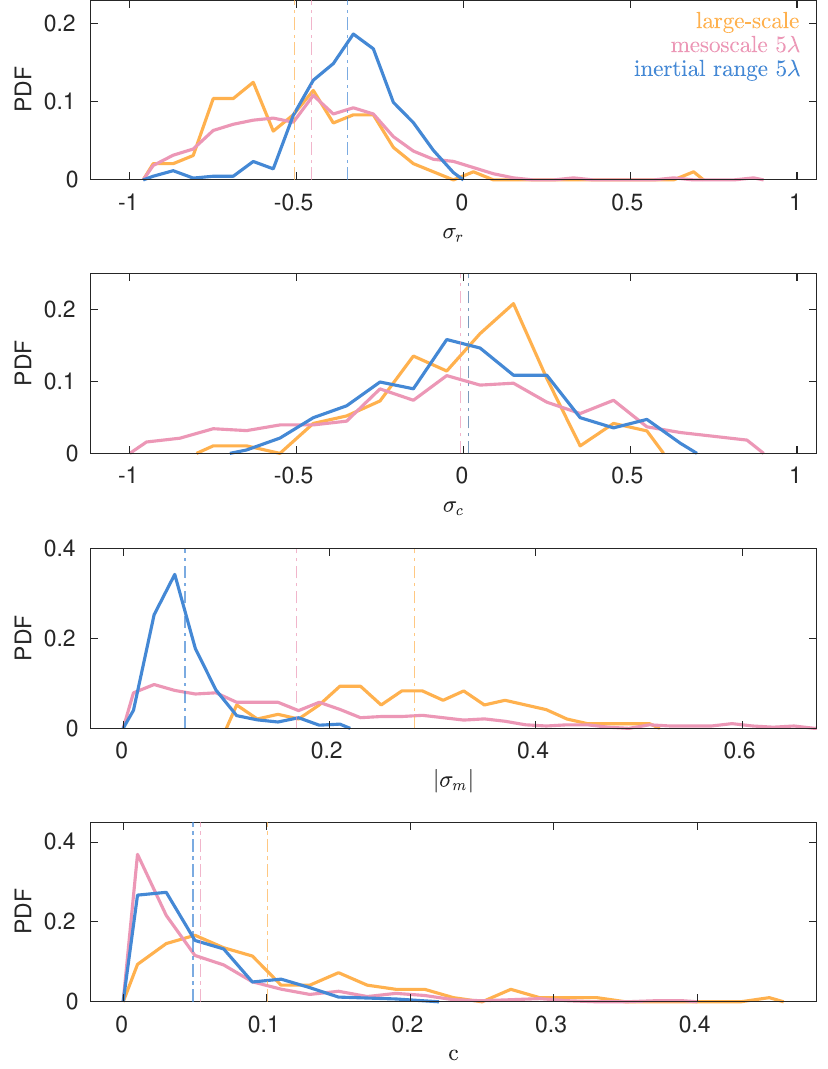}
   \caption{PDFs of the residual energy, rectified cross helicity, absolute magnetic helicity, and magnetic compressibility in the large-scale range, mesoscale range and inertial range. The latter two ranges comprise the $5\lambda$ sub-intervals. The large-scale and inertial range values are averages over the entire frequency band, while the mesoscale values are taken at the scale of the $\alpha$ maximum at mesoscales. Vertical lines mark the distribution means.}
    \label{fig:CME_sigma_histograms}
    \end{figure}

The $\sigma_\mathrm{r}$, $\sigma_\mathrm{c}$, $\sigma_\mathrm{m}$, and $c$ spectra were time-averaged in the same manner as $E_B$. As in the case of Fig.~\ref{fig:CME_alpha_histograms}, three frequency ranges were analyzed: the inertial range (here $10^{-3}$--$8\times10^{-3}$~Hz, with the upper limit set by the Nyquist frequency for the 1-min resolution ion data used to calculate $\sigma_\mathrm{r}$ and $\sigma_\mathrm{c}$), large scales (<$10^{-4}$~Hz) and mesoscales ($10^{-4}$--$10^{-3}$~Hz). The $5\lambda$ sub-intervals were analyzed for the inertial range and mesoscales but not for the large scales, as sub-intervals at the large scales were strongly affected by the cone of influence cut-out. For mesoscales, the values of the parameters defined by Eq.~\ref{eqn:sigma_r}--\ref{eqn:compressibility} were taken at the frequency associated with the least steep $\alpha$ value of each sub-interval. In the case of the other two ranges, an average was taken over the entire frequency band, as there was no singular frequency of interest in those ranges. The resulting distributions are shown in Fig.~\ref{fig:CME_sigma_histograms}. 

The top panel shows the residual energy, $\sigma_\mathrm{r}$. The mean values and standard deviations are, respectively, $-0.35$ and 0.15 for the inertial range distribution, $-0.45$ and 0.25 for the mesoscales, and $-0.51$ and 0.24 for the large scales. The large scales have the most negative $\sigma_\mathrm{r}$, as expected for the scales including the global flux rope structure. The mesoscale distribution is similar to the large-scale distribution, but with a slightly smaller absolute mean value. The mean value of the inertial range distribution is consistent with the mean value of $-0.36$ found by \citet{Good22} in ICMEs at $10^{-3}$--$10^{-2}$~Hz.

The second panel of Fig.~\ref{fig:CME_sigma_histograms} shows the rectified $\sigma_\mathrm{c}$. The mean values and standard deviations are 0.02 and 0.28 for the inertial range distribution, -0.01 and 0.42 for the mesoscales, and 0.01 and 0.25 for the large-scale. The means of the inertial and large-scale ranges are close enough to each other that the dash-dotted lines overlap; all three distributions are centered close to zero. For the large scales, the low $\sigma_\mathrm{c}$ can be interpreted as a low correlation between the global magnetic field (i.e. the flux rope rotation) and the global velocity profile, which is typically linearly declining or flat in the radial direction and zero in the non-radial directions. At smaller scales, in contrast, a time-averaged $\sigma_\mathrm{c}$ close to zero may be interpreted as a balance of sunward and antisunward Alfv\'enic fluctuations. It is notable that the $\sigma_\mathrm{c}$ mesoscale distribution has larger tails at both ends (i.e. more strongly sunward and antisunward) when compared to the other two distributions. 

The third panel of Fig.~\ref{fig:CME_sigma_histograms} shows the absolute magnetic helicity, $|\sigma_\mathrm{m}|$. The mean values and standard deviations are, respectively, 0.06 and 0.04 for the inertial range distribution, 0.17 and 0.15 for the mesoscales, and 0.28 and 0.09 for the large scales. The large-scale distribution has the largest $|\sigma_\mathrm{m}|$ values, as expected for the rotating magnetic field vector associated with the flux rope. The small values of $|\sigma_\mathrm{m}|$ in the inertial range are consistent with previous measurements \citep[e.g.][]{Smith03}. The mesoscale distribution of $|\sigma_\mathrm{m}|$ is intermediate to the other two distributions. 

Magnetic compressibility, $c$, is shown in the fourth panel of Fig.~\ref{fig:CME_sigma_histograms}. The means and standard deviations are, respectively, 0.05 and 0.04 for the inertial range distribution, 0.05 and 0.06 for the mesoscales, and 0.10 and 0.08 for the large scales. Compressibility is low in all three cases. While the mesoscale distribution is not conspicuously different to the other two distributions, it has the most sub-intervals with very low compressibility. This and the more prominent tails in the mesoscale $\sigma_\mathrm{c}$ distribution point to the mesoscales being the most Alfv\'enic of the three ranges, while the $\sigma_\mathrm{r}$ distribution suggests they have intermediate Alfv\'enicity.

\subsection{Slope uncertainties} \label{sect:uncertainties}

\begin{figure}
   \centering
   \includegraphics[width=0.49\textwidth]{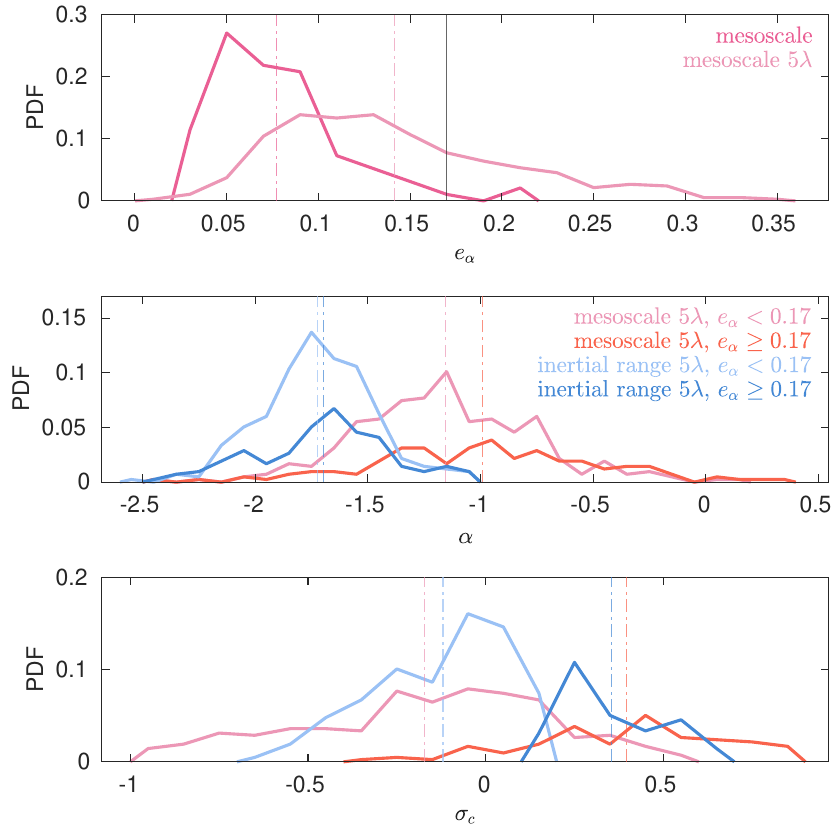}
   \caption{Distributions of uncertainties in mesoscale $\alpha$ values, $e_\alpha$, and parameters divided by larger and smaller $e_\alpha$ values. The mean values are denoted with dash-dotted lines of the same color as the distribution. The top panel depicts the $e_\alpha$ values for the full events and $5\lambda$ sub-intervals with the dividing uncertainty value shown as a black line. The two bottom panels show histograms parameter values for $5\lambda$ sub-intervals at the inertial range scale (blues) and mesoscales (reds) with distributions with smaller $e_\alpha$ values in lighter colors and larger $e_\alpha$ values in darker colors. The middle panel shows the $\alpha$ values and the bottom the cross helicity, $\sigma_\mathrm{c}$.}
    \label{fig:CME_errorsplit_histograms}
    \end{figure}

The signed $\alpha$ value has been found to increase at mesoscales. In most cases, this corresponds to the relative shallowness of a well defined linear trend in the log-log spectrum, but occasionally the spectra at mesoscales are more bumpy. In these minority latter cases, a linear fit gives $\alpha$ values with relatively large uncertainties. The distribution of the uncertainties in the largest signed $\alpha$ values at mesoscales, $e_\alpha$, for the full events and $5\lambda$ sub-intervals are shown in the top panel of Fig.~\ref{fig:CME_errorsplit_histograms}. The means and standard deviations are 0.08 and 0.04 for the full events, and 0.14 and 0.06 for the sub-intervals, respectively. Neither distribution is symmetric, with tails extending to larger $e_\alpha$ values. These tail values correspond to events with more bumpy spectra. The $5\lambda$ distribution has been used to separate $5\lambda$ sub-intervals with a well defined power law ($e_\alpha<0.17$) or bumpier spectra at mesoscales ($e_\alpha\geq0.17$); $e_\alpha=0.17$ was selected to separate the cases because it is the approximate midpoint of the distribution range, and also because it separates the tail-like part of the $e_\alpha$ distribution.

Separating events with larger $e_\alpha$ values generally has little impact on the overall results. The two most significant differences are illustrated in the middle and bottom panel of Fig.~\ref{fig:CME_errorsplit_histograms}. In these panels, the PDF normalization has been made to the total number of events across both groups, i.e., not separately for the $e_\alpha<0.17$ and $e_\alpha\geq0.17$ sub-groups. The middle panel shows the largest $\alpha$ value at mesoscales ($10^{-4}$--$10^{-3}$~Hz) and the average for the inertial range ($10^{-3}$--$10^{-2}$~Hz) for the $5\lambda$ sub-intervals similarly to the second panel of Fig.~\ref{fig:CME_alpha_histograms}, with the additional separation by $e_\alpha$. The lighter colors correspond to the $e_\alpha<0.17$ cases, with red and pink denoting mesoscales and blue shades the inertial range. It can be seen that the mean values of the mesoscale $\alpha$ distributions are at or close to $-1$, with the $e_\alpha<0.17$ distribution mean being at a marginally steeper $\alpha$ value than the $e_\alpha\geq0.17$ distribution.

The bottom panel of Fig.~\ref{fig:CME_errorsplit_histograms} shows the $5\lambda$ rectified $\sigma_\mathrm{c}$ in the mesoscale and inertial ranges similarly to Fig.~\ref{fig:CME_sigma_histograms}, with the distributions separated by $e_\alpha$. Here there is a clear difference between events with larger and smaller $e_\alpha$, with the former being almost entirely positive (i.e. net antisunward) and the latter spanning positive and negative values. The means and standard deviations of the $e_\alpha<0.17$ distributions are $-0.12$ and 0.19 for the inertial range, and $-0.17$ and 0.35 for the mesoscales. The equivalent values for the $e_\alpha\geq0.17$ distributions are 0.36 and 0.14 for the inertial range and 0.40 and 0.28 for mesoscales. This indicates that mesoscale fluctuations with a more well defined power-law scaling are more balanced in terms of a sunward versus antisunward sense of propagation, while mesoscale fluctuations with a bumpier spectrum are predominantly antisunward. It also indicates that the inertial range $\sigma_\mathrm{c}$ is determined by (or at least strongly correlated with) the mesoscale $\sigma_\mathrm{c}$. No significant differences were seen in any of the foregoing analysis when separating events according to the median $e_\alpha$ value (0.13) rather than 0.17.

\subsection{Slow wind}

\begin{figure}
   \centering
   \includegraphics[width=0.49\textwidth]{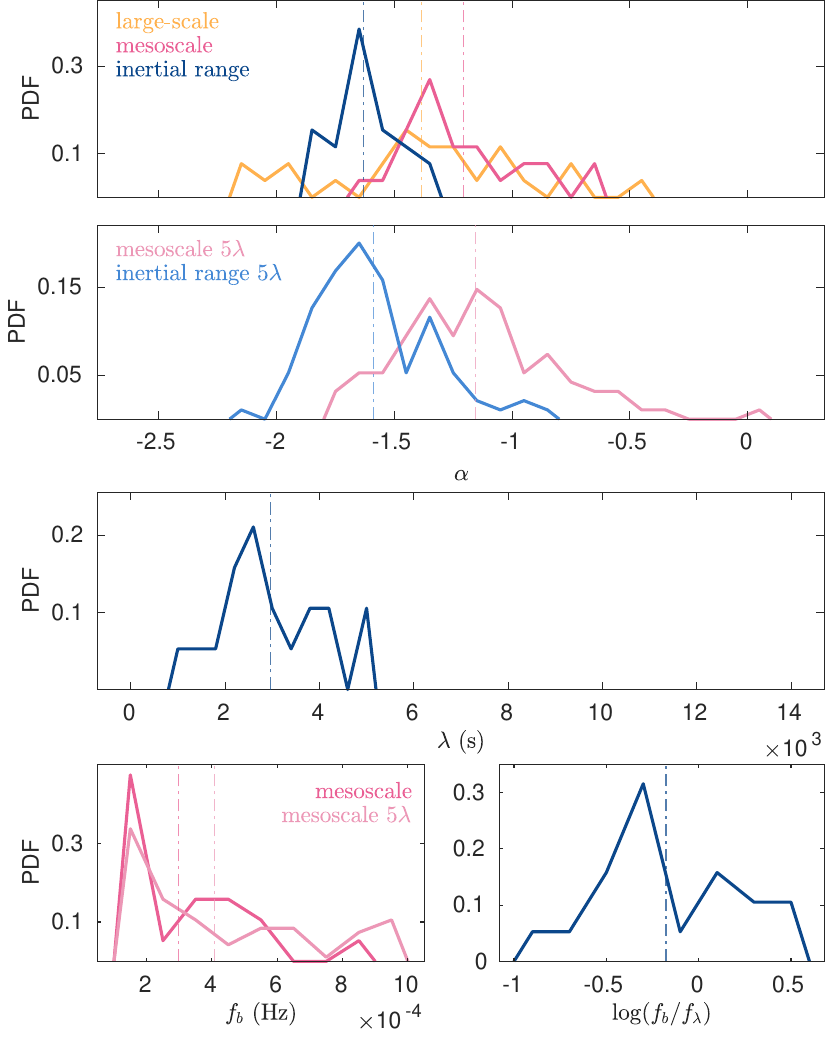}
   \caption{Histograms of $\alpha$ values, correlation lengths and frequencies of the local maximum $\alpha$ at mesoscales in an equivalent version of Fig.~\ref{fig:CME_alpha_histograms} for slow wind intervals.}
    \label{fig:SSW_alpha_histograms}
    \end{figure}

\begin{figure}
   \centering
   \includegraphics[width=0.49\textwidth]{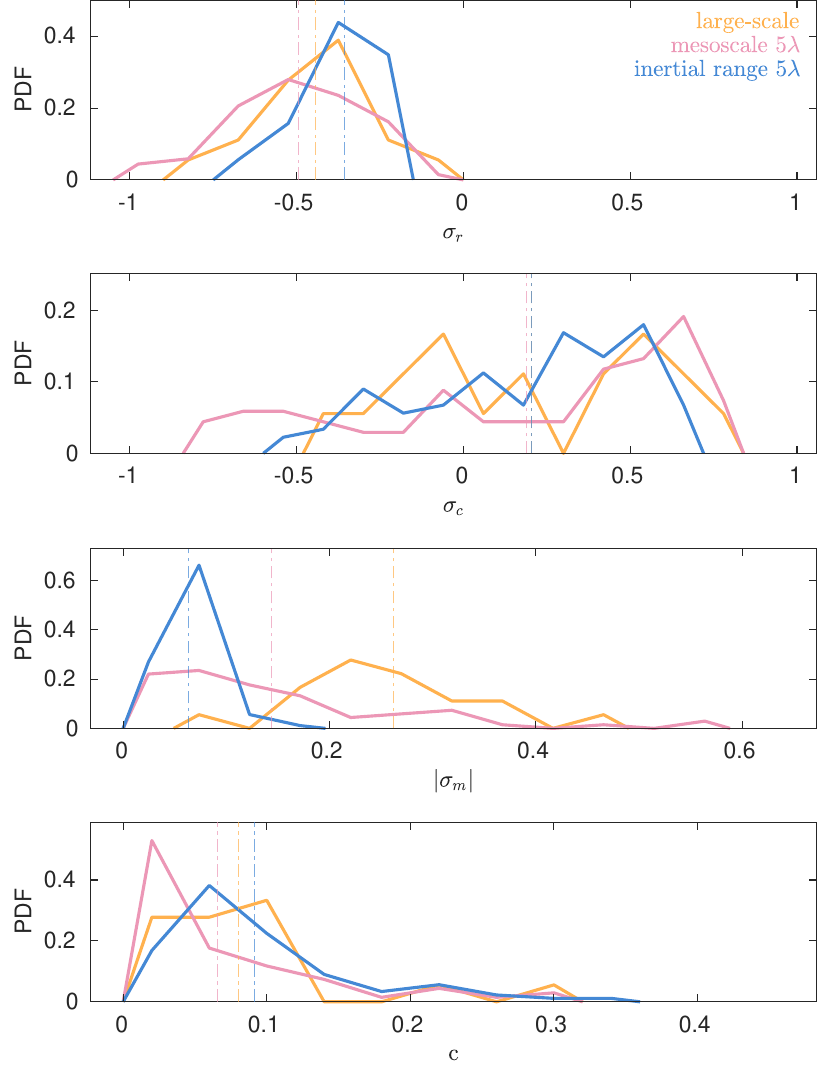}
   \caption{Histograms of the residual energy, cross helicity, magnetic helicity and magnetic compressibility in a slow wind equivalent of Fig.~\ref{fig:CME_sigma_histograms}}
    \label{fig:SSW_sigma_histograms}
    \end{figure}

A similar analysis to the one performed on ICMEs as detailed in Section~\ref{sect:individual_psds} and \ref{sect:statistical_analysis} was performed on the 26 intervals of slow wind described in Section~\ref{sect:spacecraft_data}. The results of this analysis are presented in Fig.~\ref{fig:SSW_alpha_histograms}, Fig.~\ref{fig:SSW_sigma_histograms}, and Table~\ref{tab:key_parameters}. The relatively small sample size limits the scope for statistical comparison with the ICME analysis, although the comparison is still instructive. The distributions of the $\alpha$ values at the three scales, the correlation length, and the frequencies of the largest signed $\alpha$ values at mesoscales are shown in Fig.~\ref{fig:SSW_alpha_histograms}, where the $x$-axis limits match the Fig.~\ref{fig:CME_alpha_histograms} limits to aid comparison between the two figures. The means and standard deviations of the $\alpha$ distributions in the top panel are, respectively, $-1.63$ and 0.15 for the inertial range, $-1.20$ and 0.26 for the mesoscales, and $-1.39$ and 0.42 for the large scales. The second panel shows the $5\lambda$ sub-interval equivalents for the two smaller-scale ranges, with means and standard deviations of $-1.59$ and 0.25 for the inertial range and $-1.16$ and 0.39 for the mesoscales, respectively. Both panels show that the mesoscales tend to have a spectral scaling that is less steep than the surrounding scales, as was the case with ICMEs. However, the difference between the mesoscale and large scale means is less compared to the ICME case given the shallower slopes at large scales in the slow wind intervals and, moreover, is only significant to a 10\% level in this small sample of slow wind.

The third panel depicts the correlation time. While slow wind intervals do not have a magnetic flux rope structure at large scales, a polynomial fit was subtracted from each component before calculating the correlation time so that the results are directly comparable to the ICME analysis. The mean and standard deviation of the distribution are $2.96\times10^3$~s and $1.11\times10^3$~s, respectively. We note that the distribution of correlation times without a fit subtraction did not significantly differ, consistent with the slow wind intervals having approximately constant mean fields. The correlation times in the slow intervals are typically shorter than in the ICMEs.

The bottom left panel shows the frequencies of the local maximum $\alpha$ at mesoscales, $f_b$, for the full intervals and sub-intervals, with means and standard deviations of $2.99\times10^{-4}$~Hz and $2.05\times10^{-4}$~Hz, and $4.07\times10^{-4}$~Hz and $2.86\times10^{-4}$~Hz, respectively. The full-interval distribution does not differ significantly from the equivalent ICME distribution, whereas there is a small, marginally significant (to a 4.8\% significance level) difference between the sub-interval $f_b$ means. The bottom right panel displays the distribution of the logarithmic ratio of $f_b$ and the correlation frequency. This distribution, with a mean value of $-0.18$ and standard deviation of 0.40, does show some significant difference to the equivalent ICME distribution; $f_b$ is somewhat smaller on average than $f_\lambda$ in the slow wind intervals, whereas the two frequencies are typically closer in value in the ICMEs.

Histograms of $\sigma_{\mathrm{r}}$, $\sigma_{\mathrm{c}}$, $|\sigma_{\mathrm{m}}|$, and $c$ for the slow wind intervals are shown in Fig.~\ref{fig:SSW_sigma_histograms}, in the same format as Fig.~\ref{fig:CME_sigma_histograms} for the ICMEs. The inertial range and large-scale values are means over the frequency band while the mesoscale values are the value at the frequency of the local maximum $\alpha$ at mesoscales. The inertial range and mesoscale values are $5\lambda$ sub-interval values. The top panel shows $\sigma_{\mathrm{r}}$. The mean values and standard deviations are $-0.36$ and 0.12 for the inertial range, $-0.49$ and 0.19 for the mesoscales, and $-0.44$ and 0.17 for the large-scale. Unlike in the ICME case, the mesoscales appear to be the most strongly negative in the slow wind, although the difference with the larger scales is small and not statistically significant to a 10\% level. The second panel shows $\sigma_{\mathrm{c}}$, with means and standard deviations of 0.20 and 0.32 for the inertial range, 0.24 and 0.51 for the mesoscales, and 0.20 and 0.38 for the large scales, respectively. All three distributions are weighted toward positive values, indicating the presence of the antisunward fluctuations that are expected for non-ICME solar wind. The third panel shows $|\sigma_{\mathrm{m}}|$, which does not significantly differ from the ICME results. The bottom panel shows the magnetic field compressibility, with mean and standard deviation values of 0.09 and  0.07 for the inertial range, 0.05 and 0.07 for the mesoscales, and 0.08 and 0.07 for the large scale; the mesoscales are the most incompressible. A separation by the mesoscale $\alpha$ uncertainty, $e_\alpha$, was not performed for the slow wind intervals because the total number of intervals analyzed was insufficient for a statistically valid result.

\section{Discussion} \label{sect:discussion}

The spectral features of mesoscale fluctuations inside ICME drivers observed at 1~au have been examined. We have here defined ``mesoscale'' in relation to the global flux rope at larger scales and the fully developed MHD turbulence at smaller scales, with fluctuations at the mesoscale transition having spacecraft-frame frequencies in the range $10^{-4}\ \text{Hz}\lesssim f_\mathrm{sc} \lesssim 10^{-3}\ \text{Hz}$. In the ICMEs analyzed, the spectral slope in the mesoscale range was consistently less steep than at higher and lower frequencies, with a mean value close to $-1$. This value suggests that the fluctuations at mesoscales are typically not part of the fully developed, inertial-range turbulent cascade or the global flux rope structure. In the majority of cases, the mesoscale fluctuations formed a well defined power-law spectrum spanning at least 0.7 decades in frequency, with a minority of cases having more bumpy spectra that were identified by the larger uncertainties associated with their slope values. These results were found when analyzing whole ICME intervals and sub-intervals with a duration of five correlation timescales. 

Our findings are strictly applicable to the the relatively ``simple'' case that we studied, namely ICMEs with a clear flux rope structure at large scales. Given the apparent ubiquity of the $1/f$ range at scales larger than the correlation length in the solar wind generally, we may surmise that this range would also be present in more complex ICMEs; however, this remains to be confirmed observationally. The presence of relatively shallow mesoscale slopes did not vary over time in the sample of events we analyzed, which spanned solar cycles 23 and 24.

Taken together, the $-1$ spectral index, relatively low compressibility, and rectified $\sigma_{\mathrm{c}}$ distribution centered close to zero but with prominent tails extending to $\pm1$ could indicate the presence of a balanced, non-turbulent spectrum of Alfv\'enic fluctuations at mesoscales within ICMEs. Such a balanced spectrum would contrast with the imbalance seen in the solar wind more generally (and as found in the 26 intervals of slow wind that we analyzed; Fig.~\ref{fig:SSW_sigma_histograms}), and could naturally arise from the globally closed field that typifies ICME drivers \citep{Good22}. Less commonly observed spectral bumps, which have a rectified $\sigma_{\mathrm{c}}$ distribution shifted toward positive values, could be due to sampling of ICMEs or regions within ICMEs that contain enhancements in power of antisunward-propagating waves. How such enhancements could arise is unclear, but may, for example, be related to open field within ICMEs. Furthermore, the mesoscale fluctuations we have identified in ICMEs could be an energy reservoir for the turbulence at smaller scales, akin to the $1/f$ injection range found in other solar wind types.

On the other hand, the negative residual energy at mesoscales (more negative than in the inertial range) suggests that at least some of the fluctuations have a more non-Alfv\'enic, magnetically dominated character. Such fluctuations could be an intrinsic, second-order feature of the flux rope field or more stochastic fluctuations unrelated to the flux rope; the fact that the ICMEs had a very similar  $\sigma_{\mathrm{r}}$ distribution to the slow wind intervals, which do not have a flux rope structure at global scales, would lend support to the latter possibility. \citet{Wicks13} found that fluctuations with very low residual energy ($\sigma_{\mathrm{r}}\simeq-1$) are associated with a $-1$ rather than $-5/3$ spectral index in the magnetic field PSD at low frequencies, concluding that $\sigma_{\mathrm{r}}\simeq-1$ corresponds to anti-aligned Elsasser-field vectors that are resistant to non-linear (i.e. turbulent) interactions. However, we did not find such low $\sigma_{\mathrm{r}}$ at mesoscales (mean value $-0.45$), suggesting that Elsasser anti-alignment is unlikely to be the cause (or, at most, only a minor cause) of the $-1$ mesoscale slope. The enhancement of $|\sigma_{\mathrm{m}}|$ that we have identified at mesoscales could indicate the presence of flux-rope-like magnetic structure, or could be a signature of wave polarization; it is also possible that the $|\sigma_{\mathrm{m}}|$ enhancement arises from an inverse cascade of magnetic helicity \citep{Frisch75}. The mesoscale range likely comprises a superposition of fluctuations with mixed sources and with a composition that varies between different ICMEs. In conclusion, we have found that mesoscale fluctuations in ICMEs present complex properties with a range of plausible origins. The nature and origins of these distinctive fluctuations will be the subject of further investigation.

\section{Conclusions}

We analyzed the power spectra of 142 ICMEs with clear flux rope signatures. The power spectra were computed by time-averaging wavelet spectra across the ICME intervals and shorter sub-intervals with durations of five correlation times. Spectral indices, $\alpha$, were calculated by fitting over a sliding window with a fixed width of 0.7 decades in frequency. While there was some variability between different intervals and sub-intervals, the spectra were found to have systematically less steep $\alpha$ values at mesoscales ($10^{-4}\ \text{Hz}<f_\mathrm{sc}< 10^{-3}\ \text{Hz}$), with an average value close to $-1$; this value suggests that the fluctuations at these scales are not part of the inertial-range turbulent cascade. The least steep $\alpha$ values were typically identified at scales close to the correlation length of the detrended magnetic field. The residual energy, rectified cross helicity, absolute magnetic helicity and compressibility were calculated at the large scales dominated by the flux rope, at mesoscales, and at turbulent inertial-range scales. Analysis of these parameters point to the mesoscale range in ICMEs typically comprising a mix of sunward and antisunward fluctuations, with antisunward imbalances associated with more bumpy power spectra; the fluctuations are in general magnetically dominated and may be relatively resistant to non-linear interactions, consistent with the $-1$ spectral index.

\begin{acknowledgements} \label{sec:acknowledgements}

This work was funded by a Research Council of Finland Fellowship (grants 338486, 346612 and 359914; \mbox{INERTUM}). EK acknowledges support from the Research Council of Finland's Centre of Excellence in Research of Sustainable Space (grant 352850) and Centre of Excellence in Space Resilience (grant 374096).  We thank the \textit{Wind} spacecraft instrument teams for providing the data used in this work. Wavelet spectral analysis was performed using the IRFU-Matlab analysis package available at \href{https://github.com/irfu/irfu-matlab}{https://github.com/irfu/irfu-matlab}.

\end{acknowledgements}

\bibliographystyle{aa-note}
\bibliography{references}

\end{document}